\documentclass[preprint2]{aastex}

\def\grtsim{{_ >\atop{^\sim}}}
\slugcomment{Not to appear in Nonlearned J., 45.}

\shorttitle{Effects of interactions in E/S0 paired galaxies}

\shortauthors{Hern\'andez-Toledo et al.}

\begin{document}

\title{The Effects of Interactions on the Structure and Morphology of
Elliptical/Lenticular galaxies in Pairs}
\author{H. M. Hern\'andez-Toledo \altaffilmark{1} V. Avila-Reese\altaffilmark{2} 
and 
J. R. Salazar-Contreras\altaffilmark{3}}
\affil{Instituto de Astronom\'{\i}a, UNAM, A.P. 70-264, 04510 M\'exico D. F., 
M\'exico}

\and

\author{C. J. Conselice\altaffilmark{4}}
\affil{California Institute of Technology, Pasadena, CA 91125, USA; University of 
Nottingham, UK}


\altaffiltext{1}{E-mail: hector@astroscu.unam.mx}
\altaffiltext{2}{E-mail: avila@astroscu.unam.mx}
\altaffiltext{3}{E-mail: rut@astroscu.unam.mx}
\altaffiltext{4}{E-mail: cc@astro.caltech.edu}

\begin{abstract}

We present a structural and photometric analysis of 42 elliptical/lenticular 
galaxies in (E/S0 + S) pairs observed in the $BVRI$ color bands. 
The aim of the analysis is to empirically determine the effects of interactions 
on their morphology, structure and stellar populations as seen from the
light concentration ($C$), asymmetry ($A$), and clumpiness ($S$) parameters. We
further compare these values to a control sample of 67 mostly isolated non-interacting 
E/S0 galaxies.  We find that the paired E/S0 galaxies occupy a more scattered loci 
in $CAS$ space than non-interacting E/S0's, and that the effects of interactions 
on E/S0's are minor, in contrast to disk galaxies involved in interactions. This 
suggests that observational methods for recognizing interactions at high redshift, 
such the $CAS$ methodology of Conselice (2003), would hardly detect E/S0's involved 
in interactions (related to early phases of the so called `dry-mergers'), and that 
the majority of interacting galaxies identified at high redshift must be gas dominated 
systems. We however find statistical differences in the asymmetry index when 
comparing isolated and interacting E/S0s.  In the mean, paired E/S0 
galaxies have $A$ values $2.96\pm 0.72$ times larger than the ones of non-interacting 
E/S0's. For the subset of presumably strongly interacting E/S0's, $A$ and
$S$ can be several times larger than the typical values of the 
isolated E/S0's. We show that the asymmetries are consistent with several 
internal and external morphological distortions. We conclude that the subset of 
interacting E/S0s should be dense, gas poor galaxies in systems spaning a wide 
range of interaction stages, with typical merging timescales 
$\grtsim 0.1-0.5$ Gyr. We use the observed phenomenology of this subsample to predict 
the approximate loci of `dry pre-mergers' in the $CAS$ parameter space.

\end{abstract}


\section{Introduction} \label{S1}

Understanding how elliptical/lenticular galaxies and the bulges of
spiral galaxies form and evolve is still an open question in astronomy. 
In comparison  with galactic 
disks, the properties of spheroids suggest violent or strong perturbative 
processes in their formation. There are two ways to define the formation
epoch of a spheroid: when most of its stars formed or when the stellar 
spheroid acquired its dynamical properties in violent or secular processes 
(Avila-Reese \& Firmani 1999). For the monolithic collapse mechanism 
(Eggen, Lynden-Bell \& Sandage 1962; Chiosi \& Carraro 2002),  both 
epochs coincide,  as spheroids form as a result 
of an early violent collapse and a burst of star formation (SF).

However, in the context of the popular hierarchical Cold Dark Matter 
(CDM) cosmogony, disks galaxies are generic structures and spheroids 
may form by the mergers of these disks (e.g., Kauffmann, White \& Guiderdoni 1993; 
Baugh et al. 1996), or by secular dynamical evolution   
(van den Bosch 1998; Avila-Reese \& Firmani 1999,2000). If major mergers 
occur at high redshifts when the disks are mostly gaseous, 
as appears to be the case (e.g., Conselice et al. 2003; Conselice 2006), then the 
situation is conceptually similar to the monolithic collapse.
At low redshifts most galactic disks have already transformed a large fraction 
of their gas into stars. In this case, spheroids may assemble by dissipationless
major/minor mergers and/or by disk secular evolution. But the 
picture is even more complex in the hierarchical 
cosmogony as galaxy morphology may be continuously changing (e.g., Steinmetz \& 
Navarro 2002), depending on the mass aggregation history (smooth accretion
and violent mergers) and environment. Thus, a relevant question to answer
is how the mass of early-type galaxies has been assembled.

The morphological properties of nearby elliptical/lenticular (hereafter E/S0) 
galaxies in different environments reveal important clues for 
understanding the nature of spheroids and their formation. The detection 
of fine structures, dust lanes, blue cores, nuclear disks, and peculiar kinematics 
in early-type galaxies, considered evidence of recent merging/accretion events 
(e.g., Malin \& Carter 1983; Lauer 1985; Abraham et al. 1999; Menanteau, Abraham 
\& Ellis 2001; Papovich et al. 2003; Lauer et al. 2005)
is most frequently found among galaxies in the field than in cluster members 
(Schweizer 1992; Reduzzi et al. 1996; Kuntschner et al. 2002). 
On the other hand, while the bulk of the stars in luminous cluster 
ellipticals are old and coeval (e.g., Bender et al. 1996), in low 
density environments the colors and gradient colors of ellipticals 
suggest that recent bursts of star formation occurred 
(Menanteau et al. 2001; Stanford et al. 2004; Menanteau et al.
2004). Blue clumps in early-type galaxies have also
suggested to be evidence of recent accretion episodes (Elmegreen, Elmegreen 
\& Ferguson 2005; Pasquali et al. 2006). The last authors find that the 
fraction of early-type galaxies with blue clumps increases at high redshift.

It is also possible that elliptical galaxies, at least the massive ones, 
evolve through the so-called `dry-mergers' -major mergers between galaxies 
without the presence of gas, such as ellipticals. Late interactions and 
dissipationless mergers may have a profound effect upon the population and 
internal morphology of early-type galaxies (e.g., van Dokkum 2005; Tran et 
al. 2005; Bell et al. 2005,2006), yet we know very little about how these 
effects occur, or their timescales. The results of numerical 
simulations show indeed that the outcome of the fusion of
early--type galaxies (dry merger) are anisotropic, slowly rotating, boxy 
spheroids (Naab, Khochfar \& Burkert 2006), while low mass highly rotating, disky 
spheroids are produced typically in inequal--mass spiral mergers (Naab et al. 
1999; Naab \& Burkert 2003). 

In this paper, we will study several photometric properties of 42 E/S0 galaxies in 
mixed pairs, where the companion is a spiral galaxy close in luminosity to 
the E/S0 galaxy. These early-type galaxies (at least the most
interacting) are affected gravitationally by the perturber companion, allowing 
us to explore the effects of interactions in early phases of mergers involving 
early-type galaxies  ('dry' pre-mergers). Results have shown thus far that 
nearby ellipticals in pairs and groups reveal a large variety of morphological 
and kinematic peculiarities, such as off-centering of inner versus outer isophotes, 
small gaseous disks, shells, counter-rotating cores, and the occurrence of SF in 
the recent past (Schweizer \& Seitzer 1992; Longhetti et al. 2000; Tantalo \& 
Chiosi 2004). There is also evidence that weak interactions (e.g., Thomson 1991) have 
played a significant role in determining the final structure of spheroids.

In addition to the traditional method of studying systems through their
gross photometric properties, namely color, we study the structures of these
galaxies through the concentration of stellar light ($C$), the asymmetry in 
the light distribution ($A$), and a measure of the clumpiness of light 
distributions ($S$). These three structural and morphological indices constitute
the so-called $CAS$ system, which has been proposed to distinguish galaxies at 
different stages of evolution (Conselice 2003, and references therein; see
also Lotz, Primack \& Madau 2004, who present two more
morphological parameters). Several authors are using now the $CAS$ system to 
study galaxy evolution in photometric redshift surveys (e.g., 
Moustakas et al. 2004; Conselice, Blackburne \& Papovich 2005; Cassata et al. 
2005; Menanteau et al. 2006).

We use here the $CAS$ system to address how the structures
of early-type galaxies change during an interaction.  Hern\'andez-Toledo et al. 
(2005, hereafter H2005) confirmed that the $CAS$ parameters are tracers 
of structural, morphological and SF properties of disk galaxies during an 
interaction.  To investigate how E/S0 galaxies respond to, and evolve, during 
an interaction, we use a sample of E/S0 galaxies in mixed pairs (E/S0 + S), 
which we compare to non-interacting E/S0's and galaxies involved in ongoing 
gas rich mergers (ULIRGs). We also discuss dynamical 
evolutionary mechanisms that might produce the slight structural changes 
we find in early-type systems involved in interactions.

This paper is organized as follows: Section 2 summarizes the main 
characteristics of our paired E/S0 sample, and the non-interacting 
E/S0 galaxy comparison sample. A brief description of the $CAS$ parameters 
and the methodology for measuring them in our samples is also described.  
Section 3 presents the measured $CAS$ parameters in the $R$ band for paired 
E/S0 galaxies, and compares their loci in $CAS$ space with three other samples, 
namely a reference sample of E/S0 non-interacting galaxies, an interacting/merging
ULIRG sample, and a sample of dwarf E/S0 galaxies. We also carry out a comparison 
of the $CAS$ parameters of the paired sample at different wavelenghts and explore 
the internal morphological properties of a subsample of the most interacting E/S0s. 
Our main results are discussed and interpreted in \S 4. Section 5 gives 
the main conclusions of the paper.

\section{Observational data} \label{S2}

\subsection{Mixed Morphology (E/S0 + S) Pairs}

For our study we analyze a collection of 42 {\it mixed pairs} (an 
elliptical/lenticular, E/S0, galaxy + a spiral, S, galaxy) selected from 
one of the most complete and homogeneous sample of interacting galaxies in 
the literature, the Catalog of Isolated Pairs of Galaxies in the Northern 
Hemisphere (Karachentsev 1972). Karachentsev (1972) used a strong pair isolation 
criterion in terms of the apparent angular separation and angular diameter of 
galaxies to create a list of $\sim 600$ non-merging, but mostly {\it dynamically 
bounded} pairs. Binaries selected in this way show a wide range of separations: for 
$\sim 70 \%$ of them, the projected separation does not exceed the sum 
of the diameters of the member galaxies, and only $\sim 10 \%$ of the systems 
have a separation greater than $\sim 100$ $h_{0.7}^{-1}$kpc (but less than 
$\sim 500$ $h_{0.7}^{-1}$kpc). 

For our sample of 42 mixed pairs, the mean projected separation is 
$\sim 40$ $h^{-1}_{0.7}$kpc, showing a wide range of morphological types, and 
interaction-induced features (Franco-Balderas et al. 2003, 2004, 2005).
Since our sample is restricted to isolated environments only the intrinsic properties 
of individual galaxies, and the effects of mutual interactions between pair components,
should affect the structures of these galaxies.

The sample was observed in the Johnson-Cousins $BVRI$ photometric system 
with the only criterion of availability of time and good weather conditions.
The images were acquired using two telescopes: the Observatorio Astron\'omico 
Nacional (OAN) 1.5 meter and 0.84 meter telescopes at San Pedro M\'artir, 
Baja California, in M\'exico, and under reasonable seeing conditions 
($\lesssim 1.5 \arcsec$). The data are sensitive enough to detect faint 
stellar tidal structures and at a typical size of $\sim 45$ 
arcsec/galaxy, an average of 30 resolution elements/galaxy are reasonable 
enough to avoid underestimations of the $CAS$ parameters. A more detailed 
description of the selection criteria, completeness and global optical 
emission properties can be found in Hern\'andez-Toledo et al. (1999). 
We show a mosaic of four representative E/S0+S pairs in the $B$ band to
illustrate the range of apparent separations and galaxy morphologies in 
the sample (Figure 1). 


\begin{figure*}
\vspace{14cm}
\includegraphics{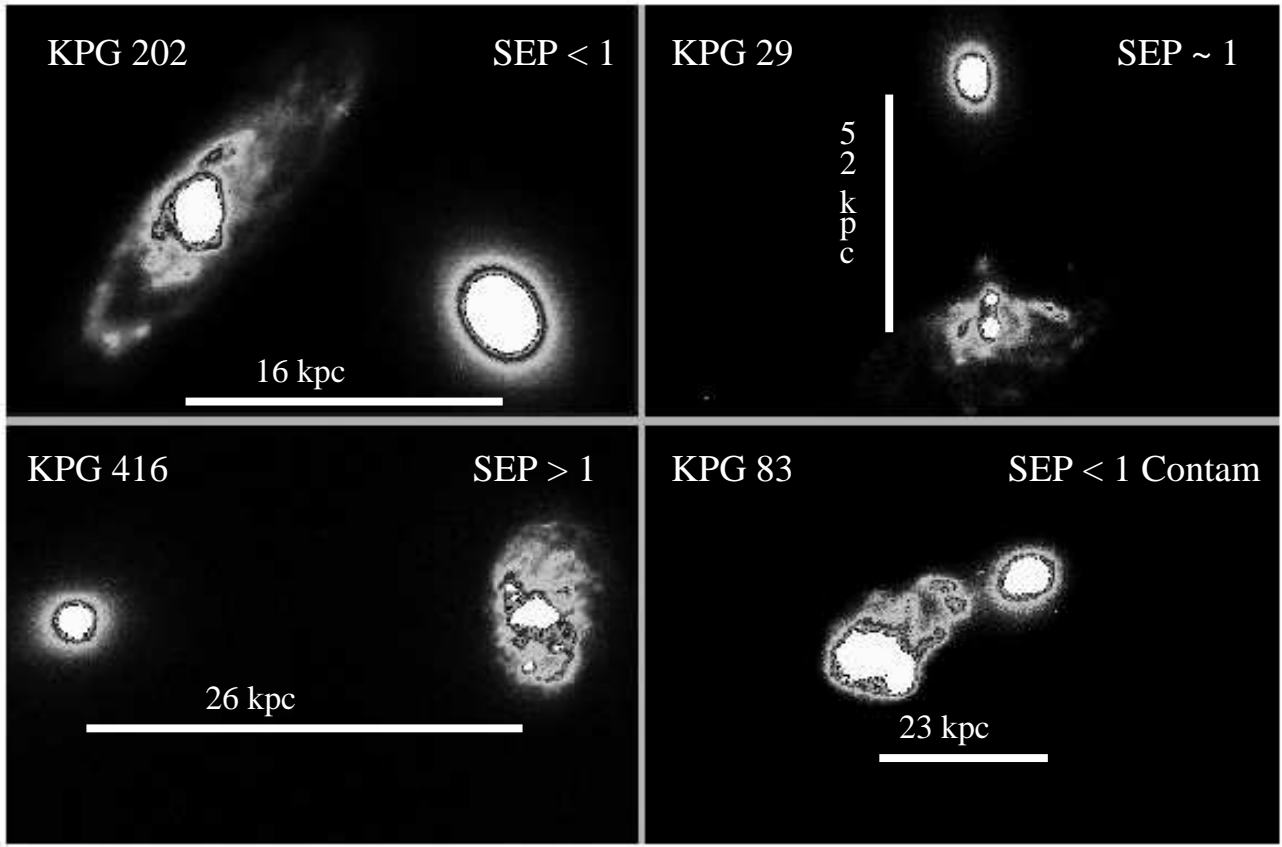}
\caption{A mosaic of $B$ band images of some reprsentative (E/S0 + S)
pairs showing a range of apparent normalized separations ($SEP$, see \S 
2.1). Upper-left panel shows KPG 202 ($SEP < 1$). Upper-right panel 
shows KPG 29 ($SEP \sim 1$). Lower-left panel shows KPG 416 ($SEP > 1$).
Lower-right panel shows KPG 83 ($SEP < 1$, where a light decontamination 
procedure is applied, see \S\S 2.1).  \label{fig1}}
\end{figure*}

\subsection{Non-interacting Galaxy Reference Sample}

As a reference sample of {\it non-interacting} E/S0 galaxies, we use images from 
three different sources: Frei et al. (1996), Colbert et al. (2001), and 
Hern\'andez-Toledo \& Salazar-Contreras (2006). The Frei et al. (1996) sample 
consists mainly of bright, high surface brightness galaxies of all morphological 
types in the field and in nearby clusters of galaxies. The images were acquired 
in two telescopes, the Lowell 1.1 meter and the Palomar 1.5 meter, with a 
typical resolution of $\sim 2$ \arcsec. From the 25 E/S0 galaxies in the 
Frei et al. sample, we have selected 18 galaxies (mostly isolated) in the $R$ band 
with no companions within a circle of 30 arcmin radius around (projected on the sky). 
  
The Colbert et al. (2001) sample is a compilation of 19 isolated and poor-group 
(non-interacting) early-type galaxies from the Third Reference Catalogue of Bright 
Galaxies (RC3, de Vaucouleurs et al. 1991). E/S0 galaxies were selected with 
no catalogued galaxies with known redshifts within a projected radius of  
$1 h_{100}^{-1}$  Mpc and a velocity of $\pm 1000 km s^{-1}$.  The $R-$band images 
were obtained at the Las Campanas 1m telescope under photometric conditions 
and a typical 1.5 arcsec seeing. 

The Hern\'andez-Toledo \& Salazar-Contreras (2006) sample is a set of 40 
isolated E/S0 galaxies extracted from the HyperLeda database, recently observed
in the Johnson-Cousins $BVRI$ photometric system at the 0.84m telescope of the 
OAN-SPM observatory with a typical resolution of $\sim 1.5$ \arcsec.  The  
basic selection criteria 
were: not belonging to any catalogued rich group and cluster of galaxies, not 
having companions in a neighborhood of 0.5 Mpc (projected on the sky) and
relative redshift lower than 600 km/s. After inspecting each galaxy in sky 
survey images for significant companions in a 30 arcsec radius (projected on 
the sky) a final sample of $30$ isolated E/S0 galaxies was considered for 
the analysis.

The final sample of non-interacting (mostly isolated) E/S0 galaxies amounts
to 67 galaxies: 18 from Frei et al. (1996), 19 from Colbert et al. (2001), and 
30 from Hern\'andez-Toledo \& Salazar-Contreras (2006). 
A typical size of $\sim 40-60$ arcsec/galaxy for all the E/S0's in this final
reference sample renders an average number of 20-30 resolution elements/galaxy, 
enough to avoid underestimations of the $CAS$ parameters (see Conselice et al. 
2000a). Homogeneity of the $CAS$ estimates for all the E/S0 galaxies in this study 
is guaranteed from the procedure described below.

\subsection{ULIRGs and dwarf Ellipticals}

As a comparison sample of galaxies involved in gas rich major-mergers we
use images of Ultra Luminous Infrared Galaxies (ULIRGs). ULIRGs are thought to 
be galaxies involved in extreme interactions, either during or after the merging 
process (Borne et al. 2000; Canalizo \& Stockton 2001). Several groups obtained 
Hubble Space Telescope 
($HST$) images of ULIRGs in the F814W (hereafter $I$) and F555W (hereafter $V$) 
bands; 66 of which have their quantitative $CAS$ structural parameters 
presented in Conselice (2003; hereafter C2003). Alternative descriptions of 
these galaxies can be 
found in Ferrah et al. (2001). To properly compare the $CAS$ values with the $R-$band 
images used in the comparison sample, a quantitative morphological $k-$correction to 
estimate the $R$-band value for each morphological index was applied (C2003). 
Considering a typical FWHM of $\sim 0.12$ \arcsec and a size of $\sim 10$ arcsec/ULIRG, 
an average number of 90 resolution elements/ULIRG avoids underestimations of their 
$CAS$ parameters.

Finally, we include in the comparative analysis the $CAS$ parameters for a sample of 19 
dwarf elliptical galaxies imaged in the WIYN telescope with the S2kB detector in the $R$ 
band. The observational details for these galaxies are described in Conselice et al. 
(2002).

\subsection{Morphological type and Magnitude distributions}

The main comparison presented in this paper is between the E/S0's in pairs
and the non-interacting E/S0 galaxy sample. To avoid inconsistencies that may 
compromise the direct comparison of these samples, we test whether their blue 
absolute magnitude and morphological type distributions are similar (Fig. 2). 
A Kolmogorov-Smirnov (K-S) test indicates that the hypothesis that the morphological 
type and magnitude distributions of both samples come from the same parental 
population can not be significantly discarded (at a significant level of 0.96 and 
0.98 respectively). Notice however that the sudden cut at the low luminosity end 
in mixed pairs is real, and has been interpreted as a morphology-density relation 
for sub-dwarf/dwarf ellipticals ($M_{B} \le -17.5$) in the field (Hern\'andez-Toledo 
et al. 1999). Early-type morphologies are equally represented in the E/S0 galaxies in 
pairs and in the reference E/S0 sample.  We conclude that the distributions 
of absolute magnitude and morphological-type are similar enough to allow a fair 
comparison between both samples in the $CAS$ space.

\subsection{The $CAS$ and SEP parameters.} \label{S2.1}

We briefly review each one of the $CAS$ parameters and discuss the reliability of 
measuring them on our samples of isolated and interacting E/SO galaxies.

{\bf Concentration of light $C$:}
The concentration index $C$ is defined as the ratio of the 
80\% to 20\% curve of growth radii ($r_{80}$, $r_{20}$), within 1.5 times the 
Petrosian inverted radius at r($\eta = 0.2$) ($r_P'$) normalized by a 
logarithm: $C = 5 \times log(r_{80\%}/r_{20\%})$. For a detailed description of 
how this parameter is computed see Bershady et al. (2000), Conselice
et al. (2002), and C2003. The 
concentration of light is related to the galaxy light (or stellar mass) 
distributions. Low (high) concentrations are expected for extended (compact) 
galaxies (Bershady et al. 2000; Graham et al. 2001; C2003). 

{\bf Asymmetry $A$:}
The asymmetry index is the number computed when a galaxy is rotated 
$180^{\circ}$ from its center and then subtracted from its pre-rotated image, 
and the summation of the intensities of the absolute value residuals of this 
subtraction are compared with the original galaxy flux. This parameter is 
also measured within $r_P'$. For a full  description see Conselice 
et al. (2000a,b). The $A$ index is sensitive to any feature that produces 
asymmetric light distributions. This includes galaxy interactions/mergers, 
large star-forming regions, and projection effects such as dust lanes
(Conselice 1997; Conselice et al. 2000a). 

{\bf Clumpiness $S$:}
Galaxies undergoing SF are very patchy and contain large amounts 
of light at high spatial frequency. To quantify this, the clumpiness index $S$ 
is defined as the ratio of the amount of light contained in high frequency 
structures to the total amount of light in the galaxy within $r_P'$ 
(C2003). The $S$ parameter, because of its morphological nature, is 
sensitive to dust lanes and inclination (C2003).

{\bf SEP parameter:} We express the apparent projected separation in paired 
galaxies ($x_{\rm 12}$) in units of the primary component diameter
($D_{\rm 25}$) where  $D_{\rm 25}$ is the 25-mag/arcsec$^2$ isophote diameter 
in the $B-$band. Thus, a quantity $SEP = x_{\rm 12}/ D_{\rm 25}$ 
is defined. We sort our (E/S0+S) sample into wide ($SEP > 1$) and close 
($SEP < 1$) pairs. Light contamination effects are expected in paired galaxies of 
similar diameters, with $SEP \lesssim 1$ or in paired galaxies with different 
diameters and $SEP << 1$.  If the light of the companion enters within the Petrosian 
radius $r_{P'}$ of a given galaxy, the observed $r_{P'}$ could be biased to a larger 
value, depending on the type of deformation induced to a ``pre-contaminated '' light 
profile. This affects the measured values of the $CAS$ parameters as shown in H2005. 
To avoid biases from this effect, the E/S0 galaxies in the closest mixed pairs were 
``decontaminated'' following a correcting procedure described below. The $CAS$ 
parameters of the ULIRG and dwarf Elliptical samples (in the $R$ band) were taken 
from Conselice (2003; 2003a).

{\bf Measurement of $CAS$ parameters.}
The measurement of the $CAS$ parameters for the E/S0 galaxies was carried out in 
several steps: 
(i) close field and overlapping stars were removed from each image; (ii) sky background 
was removed from the images by fitting a polynomial function that yielded the lowest 
residual after subtraction; (iii) the center of each galaxy was considered as the 
barycenter of the light distribution and the starting point for measurements (Conselice
2003); (iv) the $CAS$ parameters for paired E/S0's with $SEP > 1$, as well as with 
$SEP < 1$ but where the companion galaxy is beyond 2 times the $D_{\rm 25}$ of the 
primary, were estimated directly, i.e. individual components were not considered to be 
influenced by light contamination from the companion; (vi) closest pairs (those with 
$SEP < 1$ and where the companion galaxy is within 1.5 times the $D_{\rm 25}$ of the 
primary) are considered as light-contaminated by the companion. In these cases, a model 
of the pair component $a$, called {\it model a} is first built by using the task BMODEL 
in IRAF. Then we subtract this model from the original image, creating an {\it image 2}. 
From {\it image 2}, we estimated $CAS$ parameters for component $b$. This same procedure 
is applied to component $b$ of each pair to estimate $CAS$ for component $a$. BMODEL 
creates a model of the light distribution in a galaxy by taking into account all the 
information from an isophotal analysis. The resultant image, after subtracting a model galaxy 
to the original galaxy, yielded in most of the cases, traces of underlying structures. In 
such cases, these remaining features were masked or interpolated before measuring the $CAS$ 
parameters.  In a few cases, the overlapping degree makes it difficult to apply 
this procedure and thus we caution the reader by marking these pairs in our analysis.

{\bf Intrinsic ellipticity and/or inclination effects.}
Galaxies with high inclinations or axis ratios could introduce systematic 
biased trends in the values of the $CAS$ parameters (C2003), but 
usually only for disk galaxies. In the case of E/S0 galaxies the diversity 
of apparent axial ratios is therefore not expected to cause a strong effect 
in the $CAS$ estimates, in particular in the earliest types, where these 
ratios are intrinsic rather than geometrical. In any case, we have evaluated 
the influence of the apparent axial ratios on the estimated parameters by 
adopting a measure of the ``inclination to the line-of-sight'' for E/S0's as 
outlined in Paturel et al. (1997). No significant trend of $CAS$ parameters 
with axis ratio for both the non-interacting and paired samples were found. 
Notice that paired galaxies whose apparent axial ratios yield ``inclinations'' larger 
than $80^\circ$ will be marked with a skeletal circle on the following 
plots.

\section{The $CAS$ parameters of isolated and  paired E/S0 galaxies} \label{S3}

\subsection{Changes with Wavelength}\label{S3.2}

The $CAS$ parameters in the $B$, $V$, $R$ and $I$ bands for our sample 
of 42 E/S0 galaxies in mixed pairs were estimated and then analyzed for any 
relative trend. The median and average of the $CAS$ parameters for the E/S0 
components show marginal wavelength differences among the different bands. 
Both the Kolmogorov-Smirnov (K-S) test and a conventional statistics for 
measuring the significance of a difference 
of means (Student's $t-$test) between the $B$ and $I$ band $CAS$ parameters
show no significant differences. After confirming that the corresponding 
variances are not significantly different, a paired Student's $t-$test, which 
takes into account point-by-point effects in the compared samples, indicates 
that the $C$ and $A$ parameters could be marginally different at the 93\% 
level in the sense that the E/S0 components tend to be more concentrated and less 
asymmetric in the $I$ band than in the $B$ band.  Such a trend is 
more marked in the non-interacting E/S0's, at least for the 30 E/S0 galaxies 
compiled from the HyperLeda database and observed by us (see \S 2). Conselice 
(1997) and Conselice et al. (2000a) have also reported a similar difference in 
bands for the Frei+ sample. No differences with band were found in the $S$
 parameter in any of our tests. 

In the following, the $R$ photometric band is adopted to perform 
our comparative $CAS$ analysis. Table 1 shows our estimate of the $R$ band 
$CAS$ parameters for the E/S0 galaxies in mixed pairs. The $CAS$ parameters 
for the isolated E/S0 galaxies will be presented along with their 
$BVRI$ photometric properties in a forthcoming paper (Hern\'andez-Toledo 
\& Salazar-Contreras 2006).


\begin{figure}
\plotone{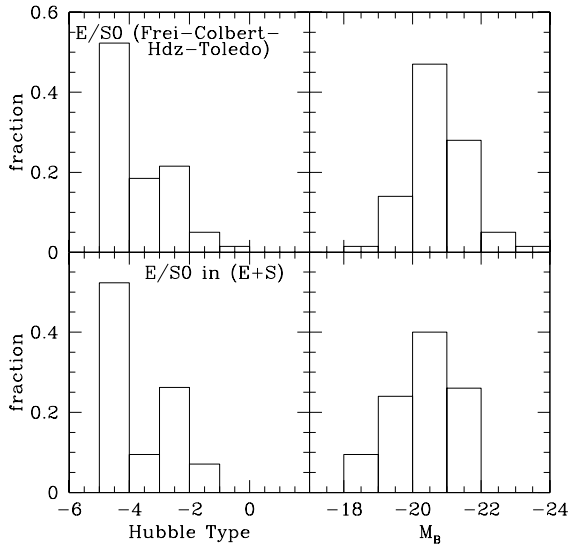}
\caption{Hubble type and blue absolute magnitude distributions of
non-interacting E/S0 (Frei-Colbert-Hdz-Toledo) and paired E/S0 galaxy samples. 
The morphological type is 
coded as E(-5), E-S0(-3), S0(-2) and S0a(0).\label{fig2}}
\end{figure}

\begin{figure}
\plotone{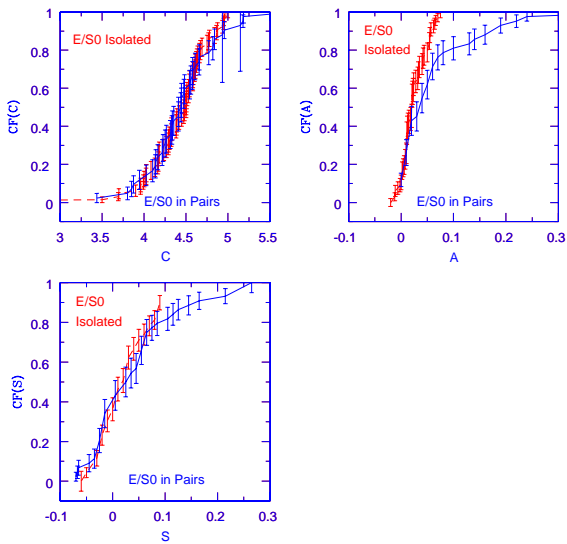}
\caption{$CAS$ cummulative distributions for the samples of paired (solid
lines) and non-interacting (dashed lines) E/S0 galaxies.\label{fig2}}
\end{figure}

\subsection{Paired vs Non-interacting E/S0 galaxies}\label{S3.3}

In this section we perform a comparative analysis of the $CAS$ parameters for
the paired and non-interacting (mostly isolated) E/S0 galaxies. Although both 
samples are in no ways complete, this comparison allows us to identify possible 
systematic differences between interacting and non-interacting E/S0 galaxies.
 We further interpret the properties of the more 
interacting E/S0s as close to the ones of early-type galaxies in a phase of 
'dry' pre-merger. Table 2 shows the median and the lower (25\%) and upper 
(75\%) quartiles of the $CAS$ parameters for E/S0 galaxies in pairs and for 
the sample of non-interacting E/S0 galaxies. Given the shape of the distributions 
of $CAS$ parameters, we adopt the medians and quartiles as more representative of the 
distributions. Table 2 also includes the results for the subsample of paired 
E/S0 galaxies in close ($SEP < 1$) pairs.

In Fig. 3 we show cumulative fractions of the $CAS$ paramaters for the
paired and non-interacting E/S0's samples in the $R$ band.
The major differences between both samples are for the asymmetry parameter.
For $\sim 75\%$ of the galaxies in each sample, $A$ is on average
$\approx 2.5$ times larger in paired galaxies than in the non-interacting, 
while for the remaining $\sim 25\%$ galaxies with the highest $A$ values, 
the difference increases even more. According to Table 2, the median $A$ value 
of paired E/S0's is $1.84$ times larger than the corresponding median of the 
non-interacting E/S0's. For the averages and their standard deviations, the
diference is $2.96 \pm 0.72$.
A Student's $t-$test confirms the statistical significance of the difference 
between the averages at the 99.5\% level, while a K-S test also shows different 
distributions at a significance level of $6.0 \times 10^{-4}$. Furthermore, 
non-parametric two-sample tests, namely two versions of the Wilcoxon Test 
(Gehan´s Generalized and Peto \& Peto), and the Logrank Test,  
reinforce the significance of this result with a probability of 0.0001. 
The statistical significant differences in the $A$ values of paired
and non-intearcting E/S0 galaxies found here imply that 
interactions do induce systematically asymmetries in the structure of 
these galaxies. However, the level of these asymmetries is typically
moderate.  
 
The median and average concentration values of the E/S0 galaxies in pairs 
are similar to those of the E/S0 non-interacting galaxies. A Student's 
$t-$test shows that the difference between both samples is not statistically 
significant 
and a K-S test also shows no different distributions at a significance 
level of $2.0 \times 10^{-3}$. The non-parametric Wilcoxon and Logrank two-sample 
tests also confirm this result. Notice however a slight difference for the subset 
with the highest $C$ values ($\sim 20\%$) in both samples: $C$ is slightly larger 
in paired E/S0's than in non-interacting E/S0's (Fig. 3).

It seems that interactions do not seriously affect the concentration of spheroids. 
However, a possibility is that the concentration could be affected, depending on the 
dynamical stage of the interaction (see \S 4.1), such that in some cases $C$ increases 
and in another ones $C$ decreases, with negligible neat differences. 

The clumpiness values of E/S0 galaxies in pairs are small ($S<0.3$), in 
spite of the interaction. The cumulative distributions of $S$ for both, the paired 
and non-interacting E/S0 samples, show that $S$ for the former is in most of cases
similar or larger than for the latter (Fig. 3); the differences in the medians
is of a factor of 1.92. However, statistical tests do not show evidence 
for any significant difference between both samples. The non-parametric two-sample 
tests also confirm this result. Thus, interactions, on average, seem not to 
systematically increase SF activity 
in E/S0 galaxies, likely because there is little gas available.  Nevertheless,
for the subsets of the $\sim 15\%$ fraction of galaxies with the largest
$S$ parameter in both samples, $S$ becomes significantly higher in the 
paired E/S0's than in the non-interacting ones, suggesting possible cross-fueling 
of gas and induced SF activity in these cases. Recent studies have shown that a 
high fraction 
of E/S0's in mixed morphological pairs show signatures of starbursts, AGN 
activity and emission excess in the 20cm continuum, which also points towards 
cross-fueling of gas from the spiral components (Domingue et al. 2003; 2005).

Figure 4 shows the $CAS$ parameters of the paired E/S0 sample versus
the apparent separation $SEP$ (left panel) and versus the ratio of the 
$I$ band absolute magnitudes of the E/S0 galaxy and its $S$ companion, 
$M_{E}/M_{S}$ (right panel). Filled circles are
for systems with apparent separation $SEP < 1$. According to Fig. 4, there is
no significant correlation of the $CAS$ parameters with $SEP$ or
the $M_{E}/M_{S}$ ratio. In more detail, one sees that most galaxies 
with the highest values of $A$ have $SEP < 1$ and$/$or $M_{E}/M_{S}\approx 1$.

\begin{figure}
\plotone{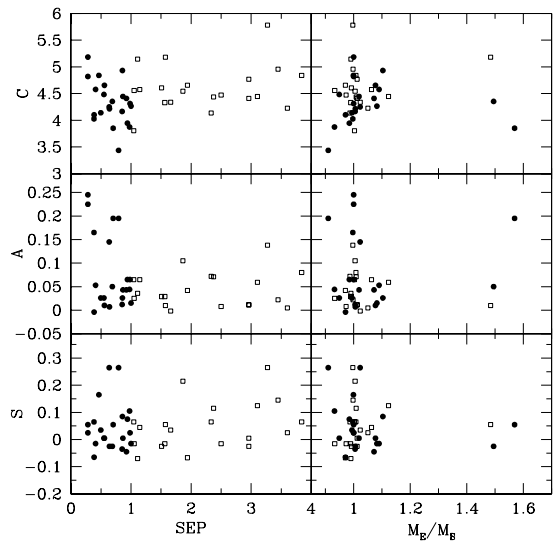}
\caption{$C$, $A$, and $S$ parameters vs $SEP$ (left panel) and vs the 
$M_{E}/M_{S}$ ratio (right panel) for the paired E/S0 galaxy sample. Filled 
circles and squares label paired galaxies with separation $SEP < 1$ and $SEP > 1$, 
respectively. $M_{E}/M_{S}$ is the ratio of the E/S0 $I$ band 
absolute magnitude to the spiral companion $I$ band absolute magnitude in each pair.
 \label{fig3}}
\end{figure}

\subsection{Loci of Mixed Pairs in the Structural $CAS$ Parameter Space}

In Fig. 5 the loci of paired E/S0 galaxies in the $R-$band $A-C$, $A-S$, and 
$S-C$ planes of the $CAS$ space are shown. The data are sorted into two 
groups:  close ($SEP < 1$, circles) and wide ($SEP > 1$, squares) paired
galaxies. For reference, the mean projected separations of the former and 
latter subsets are $<x_{1,2}> \approx 24 h^{-1}_{0.7}$kpc and 
$<x_{1,2}> \approx 75 h^{-1}_{0.7}$kpc, respectively. The solid circles 
represent decontaminated $CAS$ values for the 
eight closest pairs after applying the procedure described in \S 2.5 and H2005. 
The lenticulars are marked with a small horizontal line crossing 
the corresponding symbol. Galaxies with ``inclinations''
greater than $80^{\circ}$ are marked with a cross. For comparison, the $CAS$ 
values of the non-interacting sample are plotted as crosses. Additional 
error bars correspond to the average and $1\sigma$ dispersion of the $CAS$ 
parameters of ULIRGs and dwarf E/S0's, from Conselice (2003; 2003a).

 \begin{figure*}
\epsscale{2}
\plotone{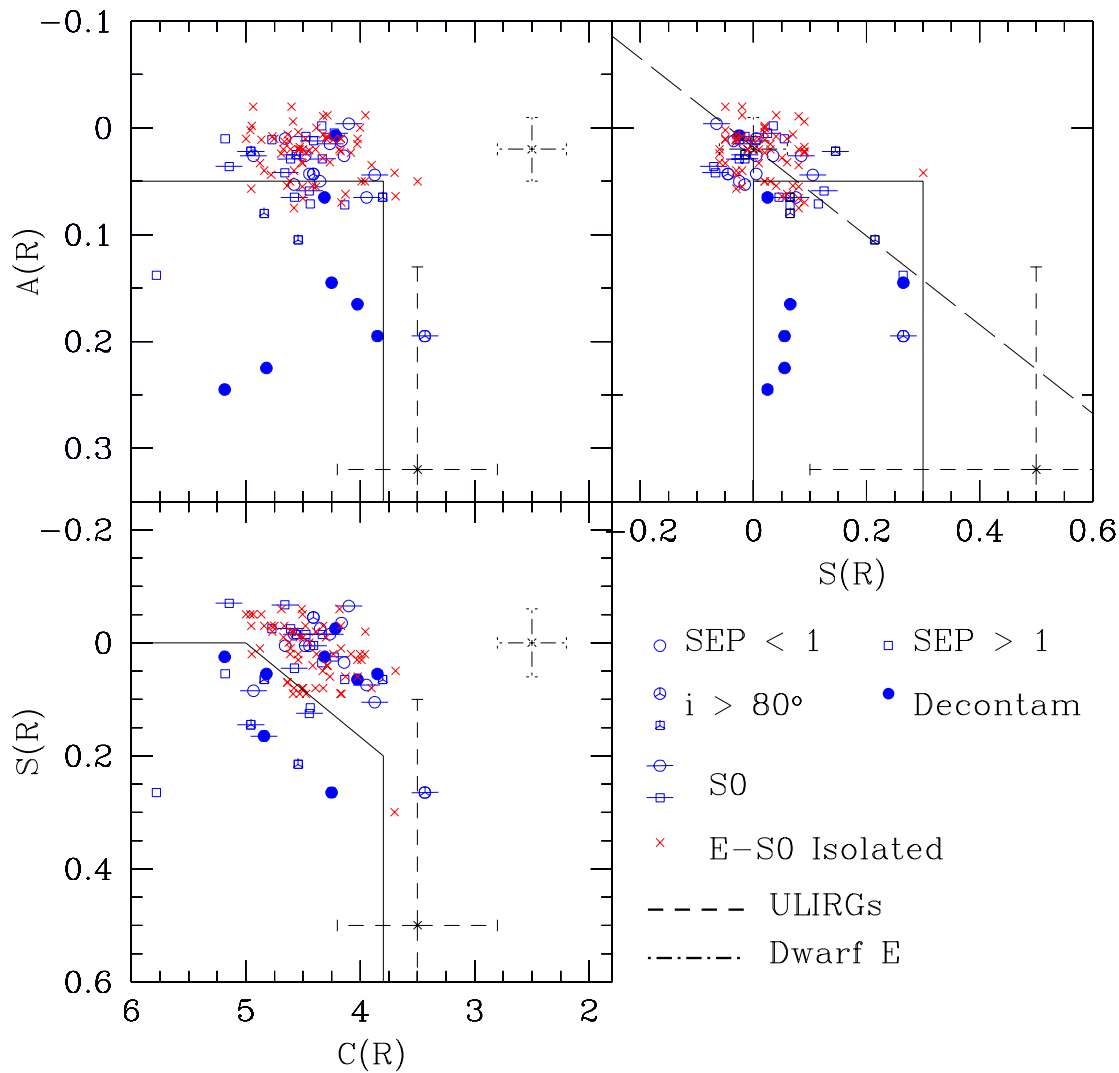}
\caption{Loci of the E/S0 paired and non-intracting galaxies in the $R-$band 
$CAS$ planes. The different symbols are explained inside the figure. 
Also shown are the average and $1\sigma$ dispersion loci of samples of ULIR and 
dwarf E/S0 galaxies (dot-dashed and short-dashed error bars respectively, see 
Conselice 2003). Regions inside 
the solid lines show the $CAS$ loci for presumably strongly interacting E/S0 
galaxies in pairs (potential pre-merger precursors of bright ellipticals). 
\label{fig3}}
\end{figure*}

The loci in the $CAS$ space of most E/S0's in pairs coincide roughly 
with that of the non-interacting E/S0 galaxies, but the former are much more 
scattered in this space than the latter (Fig. 5).
For a subset of the paired E/S0's with the highest asymmetries, a weak trend 
is seen such that as $A$ is larger, $C$ is larger. Notice that most of
these galaxies had to be decontaminated by the presence of a close companion  
(see \S 2.5 and H2005).

In the $S-C$ plane, a trend of increasing $S$ as $C$ decreases
is observed for the non-interacting E/S0's (Fig. 5). The paired E/S0's also
show this tendency although with more scatter and with a subset
shifted to high $S$ values. Most of the galaxies of this subset are S0s.
Furthermore, in the $A-S$ plane, the weak trend of increasing $A$ as $S$ increases 
is seen for both non-interacting and paired galaxies. For the paired E/S0's with 
the highest $A$ values (mainly E's), the values of $S$ are not as high as
one would expect by extrapolating the trend of the rest of the galaxy sample.

A view of the $CAS$ planes reveals a subsample of E/S0s in pairs
whose behavior clearly departs from the rest of the paired E/S0s (and
of the non-interacting ones). These galaxies have typically $A$ and $S$ 
values larger than the average or median (see also \S 3.2 and Fig. 4).
The regions enclosed in solid lines in Fig. 5 are an attempt to represent
the approximate loci of these galaxies in the $CAS$ planes.
In \S 4.2 we explore more closely the morphological appearance of
these galaxies and show that they have features consistent with
high values of $A$ and $S$ parameters.

Finally, we can see in Fig. 5 that the $CAS$ parameters of interacting
E/S0's are far from those of the ULIRG and dwarf E samples. ULIRGs imply 
mergers with a significant fraction of gas, more easily detected through
their larger asymmetry and clumpiness (e.g., C2003; Lotz et al. 2004).

\subsection{$CAS$ parameters and global galaxy properties}

Figure 6 shows the $CAS$ parameters vs. corrected absolute $B-$band 
magnitude M$_B$, $(B-V)$ color index, central velocity dispersion 
$\sigma_0$ in logarithmic scale, and the continuum 20cm of the 
luminosity L$_{\rm 20cm}$ also in logarithmic scale, for paired E/S0 
(open symbols) and non-interacting E/S0  sample (crosses) galaxies. 
Color data for the paired galaxies were taken from our photometric studies 
(Franco-Balderas et al. 2003, 2004, 2005), while absolute magnitudes and 
velocity dispersions were taken from the HyperLeDa 
database\footnote{http://leda.univ-lyon1.fr}. Magnitudes, colors, and velocity 
dispersions for the reference sample were also taken from HyperLeDa database, 
while $L_{20cm}$ luminosity was estimated from data compiled from the NRAO/VLA 
Sky Survey at 20cm (NVSS) for both samples.

For both the paired and non-interacting E/S0's samples there are no significant 
correlations between $CAS$ parameters and  M$_B$, $(B-V)$, $\sigma_0$, and 
L$_{\rm 20cm}$. The loci of paired E/S0's galaxies in all the diagrams of Fig. 6 
shows again more scatter than that of non-interacting galaxies but clear and 
systematic differences are not appreciated, except for the $CAS$ vs 
$(B-V)$ color plots. While no non-interacting E/S0 galaxies have $(B-V) \lesssim 
0.8$, a high fraction of the most asymmetric and clumped paired E/S0 galaxies 
are bluer than $(B-V)= 0.8$. In general, paired E/S0's contain a larger
distribution of colors ($\sigma_{B-V} = 0.125$) than the non-interacting E/S0's 
($\sigma_{B-V} = 0.060$), and on average the former are slightly bluer than 
the latter ($<(B-V)>= 0.84$, and $<(B-V)>= 0.89$, respectively), showing 
that SF in the recent past could likely be induced to some level 
by gas accreted from the spiral companion. This conclusion is somehow reinforced 
by the surprising NVSS radio detection fraction found in the E/S0 components in our 
sample of mixed pairs (see last panel of Figure 6 and Domingue et al. 2005) 
suggesting that the gas giving rise to this effect may be acquired from the gas 
rich companion.

\begin{figure*}
\epsscale{2}
\plotone{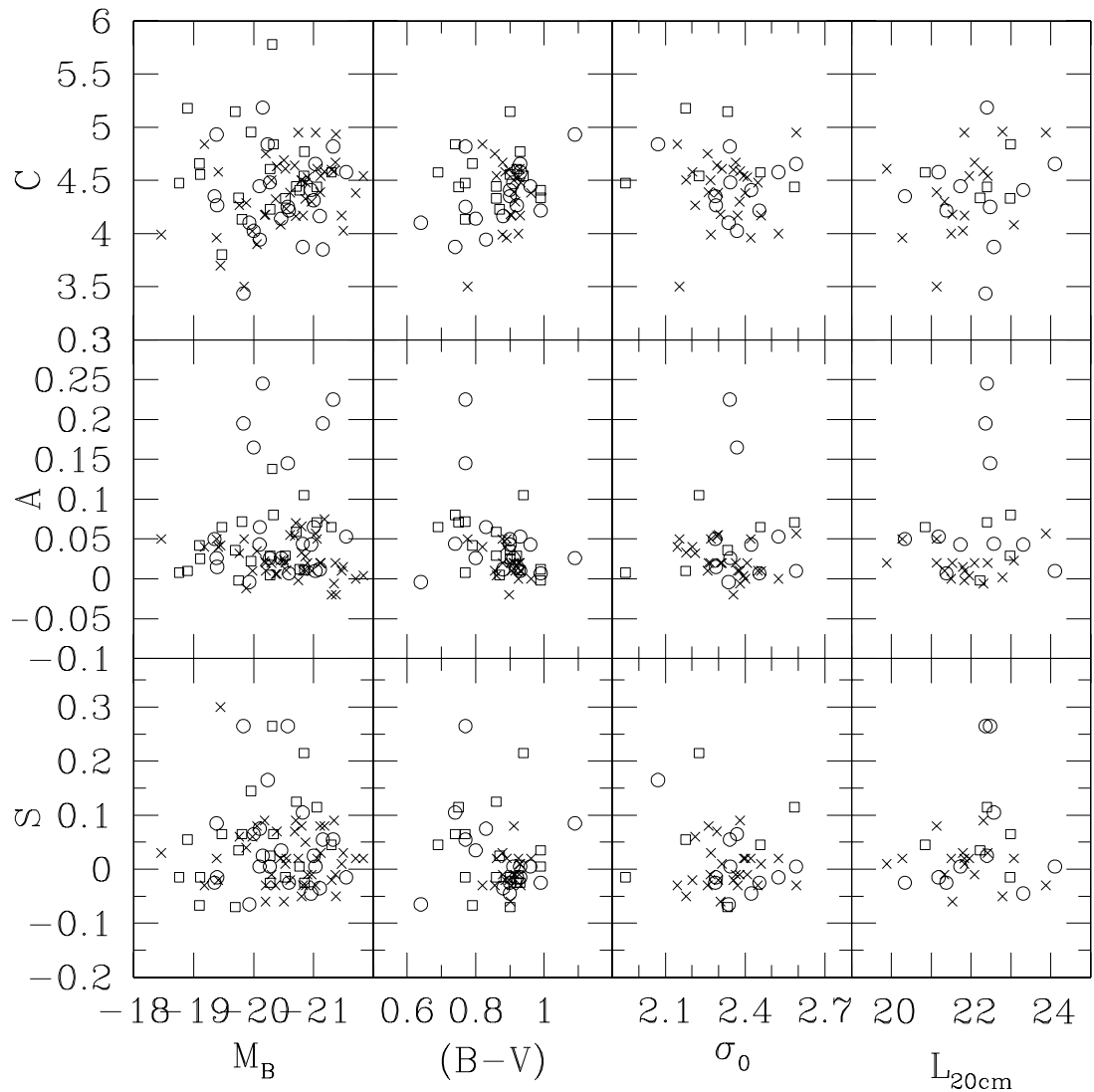}
\caption{$R$-band $CAS$ parameters vs blue absolute magnitude $M_{B}$, 
integrated corrected $(B-V)$ color, logarithmic central $\sigma_{o}$ velocity 
dispersion, and logarithmic 20cm luminosity (the latter from the NVSS survey). 
Non-interacting E/S0 galaxies (crosses) as well as close ($SEP < 1$) E/S0 paired 
galaxies (open circles) and wide ($SEP > 1$) paired galaxies (open squares) are 
shown. \label{fig4}}
\end{figure*}

\section{Discussion} \label{S4}

\subsection{The effects of interaction on E/S0 galaxies}

If E/S0 galaxies in mixed pairs originate in a similar way as the 
isolated E/S0 ones and therefore have similar initial $CAS$ parameters, 
then it is important to understand how, as a consequence of the 
interaction, $A$ becomes larger at least for a subset of the paired E/S0's.
These galaxies may well represent the properties of early-type galaxies
in the phase of ``dry'' pre-merging. 

Numerical simulations of E+E interpenetrating encounters have shown that 
$m=1$ deformations are common when the pericenter is comparable or less 
than the characteristic size of the system. At large interacting distances 
the $m=1$ term disappears and the bi-symmetric ($m=2$) term dominates 
(Combes et al. 1995). The signature of $m=1$ deformations include 
off-centering of inner isophotes compared to the outer ones, which on its 
own will cause an asymmetry signature. The numerical simulations 
also show that the intermediate and outer envelopes remain off-centered or 
asymmetric for only a few $10^8$ years after the pericenter passage (Combes 
et al. 1995; see also Aguilar \& White 1985). Although very noisy, a loose 
trend of decreasing $A$ as $SEP$ is larger (Fig. 4) is observed, possibly 
reflecting the short time scales of $m=1$ deformations. 

It could be expected that collisionless spheroids expand during 
interactions as orbital energy of the perturber is transferred to 
internal energy of the spheroid (e.g., Aguilar \& White 1985; 
Namboodiri 1995; Vergne \& Muzzio 1995; Evstigneeva, Reshetnikov \& Sotnikova
2002). However, as all these authors show, the process is much more complex and 
depends on several initial and physical conditions of the pair, for example 
the mass ratio of the intervening galaxies, orbital parameters, strength of 
the interaction, and initial structure of the components. Furthermore, the shape 
of the surface density profile is expected to change during the interaction, 
reaching its equilibrium state after several crossing times. In this scenario, 
a unique global parameter like $C$ will not be sensitive enough to describe 
all (global and local) structural changes and transient processes associated 
with the interaction.   

We have found that the concentration index tends to be the same in paired 
E/S0 galaxies as in the non-interacting E/S0's, but with a larger
dispersion for the E/S0 sample, showing this the diversity of dynamical 
evolutionary stages through which the pairs are observed. This is consistent 
with results from several numerical simulations. In these simulations there 
are cases when the galaxy's surface density profile becomes more compact. 
Simulations of spheroids with density profiles given by the de Vaucouleurs models 
that are tidally perturbed by encounters show that strong collisions 
produce a final shrinkage in the effective radius and a brightening in the 
effective surface brightness, whereas weak collisions have the opposite effect 
(Aguilar \& White 1985). However, during early stages of the interaction the 
profiles flatten externally and deform. Aguilar \& White (1985) conclude that the 
effects of tidal encounters in the luminosity profiles can only be recognized 
immediately after a close passage. 

On the other hand, we caution that the concentration index does not describe 
in detail the diversity of luminosity distributions. It could be that in some 
cases the expansion is efficient only in the outer, less bounded regions, in 
such a way that $r_{80\%}$ increases but $r_{20\%}$ remains almost constant or 
even decreases (e.g., see Namboodiri 1995 for head-on E+E simulations). 
Therefore, the concentration index will ``formally'' increase in spite of 
the profile shows an outer expansion.

The $CAS$ parameters of E/S0's do not change much during an interaction, in
comparison to interacting disk galaxies.  The structures of spheroid dominated 
galaxies also show less distortion during a merger than a pure disk system in
N-body models (Conselice 2006). The small observed change of
E/S0 galaxy properties with interaction agrees with the fact that early-type 
galaxies are typically dense, concentrated, dynamically hot structures with low 
gas fractions which are slowly rotating. Furthermore, the interacting S0 galaxies have 
$CAS$ values closer to the non-interacting ellipticals rather than to the interacting 
spirals, implying thick, hot and gas-poor disks, which do not strongly react to the 
interaction.

Finally, we consider the possibility that the differences seen between E/S0s in 
mixed pairs and non-interacting E/S0's are due to differences in their formation 
processes rather than to the interaction as disscused above. The E/S0's in pairs 
could originate from disks (Sa-Sb galaxies) by secular mechanisms amplified 
by interactions. However, the interacting spirals have $CAS$ parameters (see
H2005) that are very different than those in the paired E/S0 galaxies; unless 
the transition is very quick, which is unlikely, this result does not support 
the secular mechanism of E/S0 formation in pairs.  In the secular scenario
it is also expected that the formed pseudo-spheroid will have a significant
level of rotation. Numerical simulations show that the level of asymmetry, including 
off-centering, in spheroids is much more pronounced if they are rotating 
(Combes et al. 1995). Thus, the relative low level of asymmetry measured in the 
paired E/S0 galaxies suggests that the rotation of these galaxies is probably slow, 
and thus do not originate from latter Hubble types.

\subsection{Identifying Strong Interactions in E/S0's with $CAS$ Parameters}

Conselice (2003), and more recently H2005, have inferred statistical criteria 
for identifying galaxy major pre-mergers based on their $CAS$ parameters.
Thus, interacting disk galaxies can be identified by their $CAS$ values in 
an automated way in high-redshift samples. It would be useful to have the 
analogous criteria for finding pre-merging E/S0's, and whether they
meet the criteria for interaction established in C2003,
and Conselice et al. (2000a,b).

For our sample of E/S0 galaxies in mixed pairs, we have found some
evidence of systematically higher asymmetries than in the non-interacting
E/S0 galaxies. In particular, for the apparently most interacting E/S0 galaxies,
the $A$ parameter can be several times larger than the typical $A$ value of
non-interacting early-type galaxies. However, even in these cases,
the value of $A$ is not larger than 0.35, which was suggested as a 
rough criterion of ongoing merger or interaction for disk galaxies 
(C2003; H2005). This indicates that the structures
of E/S0's are very robust to changes due to interactions because of their 
high stellar mass density and likely dark matter concentrations, and/or a 
lack of gas content and slow rotation. This is an indication that mergers 
between evolved galaxies without significant gas, the so-called 
`dry mergers', would be underestimated in merger studies that rely on morphology.  
Studies that use pair techniques to find mergers (e.g., Bundy et al. 2004; Lin et 
al. 2004) might be necessary to fully account for these systems. 
The introduction of more morphological indicators, as the Gini coefficient and 
$M_{20}$, seem to be also of utility (Lotz et al. 2004,2006).
As a result, it is very likely that the progenitors of mergers found at 
high redshift are gas-rich disk galaxies rather than ellipticals
(Conselice et al. 2003a,b).

Based on our results, we can develop a loose structural criteria for 
finding early-type galaxies perturbed by a companion of nearly similar
luminosity (mass). In our non-interacting sample of E/S0's, approximately 
$80\%$ have asymmetries smaller than 0.05, while for our paired E/S0 sample,  
$\sim 45\%$ of the galaxies have $A$ values smaller than 0.05. All of 
these galaxies are in pairs with morphological  
indices that reflect weak evidence of tidal distortions. One may therefore
roughly define $A=0.05$ as the transition value from non-interacting 
to interacting E/S0 galaxies, although reaching this level of accuracy
in the asymmetry index at high redshift is difficult (Conselice et al. 2000a).

Since there is some trend of $A$ with $S$ for both the paired and non-interacting
E/S0 galaxies (Fig. 5), it is important to explore the deviations of
paired from non-interacting galaxies in the $A-S$ plane. We define the
normalized deviation $\sigma_{AS}$ as  $(A-\overline{A}_{\rm n-i})/\sigma$,
where the numerator is the residual of the observed $A$ value of a paired
E/S0 galaxy with respect to a linear bisector fitting to the sample of 
non-interacting E/S0 galaxies in the $A-S$ plane, $\overline{A}_{\rm n-i}$, 
and $\sigma$ is the variance of the fitting.  The deviations of paired E/S0's 
from the non-interacting ones in the $A-S$ plane can be as large as 
$\sim 10\sigma$.  In Fig. 7 we plot $\sigma_{AS}$ vs $A$ for our E/S0 pair
sample.   Most galaxies have $\sigma_{AS}>0$, and  $\sigma_{AS}$ increases
as $A$ is higher, i.e. the $S$ parameter of paired E/S0's increases with $A$
at a slower rate than in the case of the non-interacting E/S0's.
This is evidence that the SF does not increase significantly even for galaxies 
with relatively large asymmetries, i.e. galaxies with significant interaction. 

In Fig. 7 galaxies are coded according to the interaction indices introduced
by Karachentsev (1972). The index assignment is based on a visual inspection 
of morphological characteristics of the system components. The $AT$ index is 
designated for pairs with components in a common luminous halo with an amorphous, 
shredded, or asymmetric structure. $LI$ pairs show evidence of tidal bridges, 
tails or both in discernible components. $DI$ pairs show evidence of structural 
distortion in one or both of the separated components. Finally, a $NI$ index 
is introduced for wide pairs with no obvious morphological distortions. The 
order $AT-LI-DI-NI$ has been suggested (Hern\'andez-Toledo et al. 2001) as a 
sequence from strongest to weakest evidence for tidal distortion in the pair 
or, alternatively, most to least dynamically evolved. 

\begin{figure}
\vspace{4.7cm}
\includegraphics{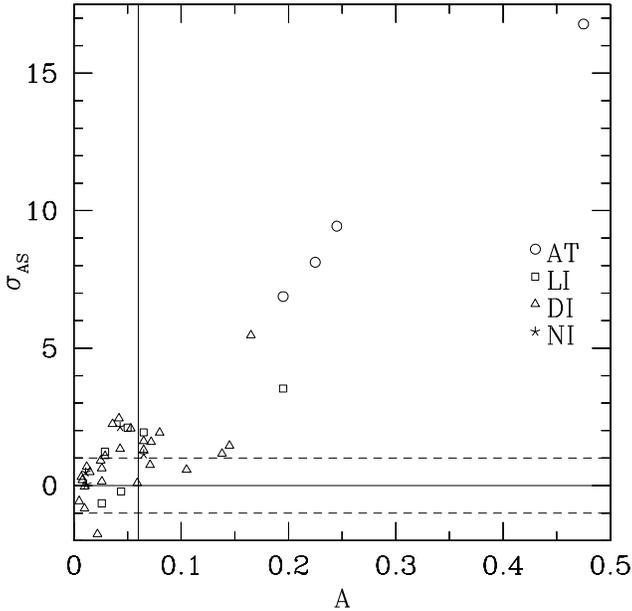}
\caption{Deviations in $\sigma$ units of paired E/S0 galaxies from the bi-sector
lineal correlation of the non-interacting E/S0 sample in the $A-S$ diagram vs 
$A$. Two strong interaction criteria are illustrated: $A>0.05$ (right vertical 
solid line) and $|A-\overline{A}_{\rm isol}|>1\sigma$ (out from the dashed line 
band). Symbols indicate the interaction index of the pair to which the E/S0 belongs
(see legends in the figure). The order from AT, LI, DI to NI 
characterizes a sequence from strongest to weakest evidence of dynamical 
interaction. The few paired E/S0's that lie below both
criteria have indeed indices (DI and NI) that denote weak evidence
of dynamical interaction. \label{fig5}}
\end{figure}

From Fig. 7 one infers that a statistical criterion for finding presumably 
interacting E/S0 galaxies might be $A > 0.05$ and $\sigma_{AS} > 1$, the former being 
the stronger criterion. The galaxies below these criteria belong to pairs with weak 
evidence of interaction (interaction indexes $DI$ and $NI$, and $SEP > 1$). 
From an inspection of Fig. 5, one might complement the above criteria of presumably 
interacting early-type galaxies with the extra conditions $C > 3.8$ (valid also for 
the non-interacting E/S0's), and $0.86 -0.17C < S < 0.3$ (see regions enclosed by 
solid lines in Fig. 5).
  
We further analyze the subsample of E/S0 galaxies with $A\ge 0.05$ to look 
for the presence of any internal/external morphological signatures. Notice 
that some of them are identified as E/S0s in close pairs ($SEP<1$) and their light 
had to be decontaminated by the presence of the spiral companion (see \S 2.5). 

Observations of close encounters involving an early-type galaxy reveal that 
morphological signatures of interactions are often weak; tidal tails are broad and of 
low surface brightness. Our ability to detect low surface brightness features is limited 
by the extent of the features, confusion with neighboring objects, contrast of the 
image and other effects. We proceed to look for the presence of tidal external 
signatures by analyzing our imges via unsharp masking to the complete image frame 
(including the spiral companion). Next, the light distribution in each E/S0 companion 
is modelled by using the GALFIT package (Peng et al. 2002). This elliptical model 
is then subtracted from the original image to look for internal structure in a 
residual image. Additionally, a $(B-I)$ color-index map is obtained to complement 
our search of tidal/localized features. Figure 10 illustrate the results 
of our image analysis for 3 representative galaxies from the subsample of the most 
interacting E/S0 galaxies with $A\ge 0.05$.

For the subsample of the most interacting E/S0 galaxies with $A\ge 0.05$, 5 E/S0s show clear 
central elongated or disky structures, 6 E/S0s show 1 or 2 localized clumps along the face of 
the galaxy, and 3 E/S0s show evidence of shell-like structures. These results are consistent 
with recent studies of local isolated ellipticals surveyed for internal structures (Colbert et al. 
2001; Reda et al. 2004) and elliptical galaxies in the Ultra Deep Field (Elmegreen et al. 2005; 
Pasquali et al. 2006). Whether these structures are produced by the interaction or not, the presence 
of central disks (in a few cases) that might react to the interaction as well as the presence of 
localized clumps out of the central regions could lead to a formal increase of the asymmetry 
parameter. This is further confirmed by viewing the residual map from the asymmetry computation. 
The cases with localized blue clumps are probably related to enhanced SF in a particular 
region of the galaxy. However, this cannot be associated with a global increase of the SF 
activity, as typically happens in interacting spiral galaxies, where the $S$ parameter is 
systematically enhanced overall the galaxy image. Thus, from our preliminary image analysis of 
the subset of presumably interacting E/S0 galaxies, we conclude that the observed 
internal/external structures deviced in these galaxies are consistent with their observed 
levels of asymmetry and clumpiness.

If the above morphological features are interpreted as signatures of past and ongoing 
interactions in these pairs, it is important then to have an estimate of their 
timescale of merging. Following Patton et al. (2000), we assume that the timescale 
for merging of inequal mass galaxies can be approximated by the dynamical 
friction timescale. We use the mean values of the $B-$band magnitude, projected 
separation (it gives a lower limit to the physical separation $R_p$), and relative 
velocity of the subsample of 17 E/S0's with $A>0.05$: 
$-20.2^m$, 22 kpc, and 125 km/s, respectively. Thus, assuming $M_s/L_B=8$ for these 
E/S0 galaxies, the average dynamical friction (merging) timescale of the subsample 
is $\grtsim 0.1$ Gyr, which is relatively short.  For nearly-equal
mass pairs, an alternative estimate for the merger timescale is provided by just
the circular orbit timescale, $t_{\rm orb}\approx 2\pi R_p/V_c$, where
$V_c$ is the galaxy circular velocity, related to the effective velocity dispersion
by $V_c\approx \sqrt{2}\sigma_v$. This effective velocity dispersion is smaller than 
the central one, $\sigma_0$. Assuming an effective velocity disspersion 
$\sigma_v\sim 180$km/s for the subsample of highly assymetric E/S0's, we estimate 
an average merger timescale of $\grtsim 0.5$ Gyr. 

\begin{figure*}
\epsscale{2}
\plotone{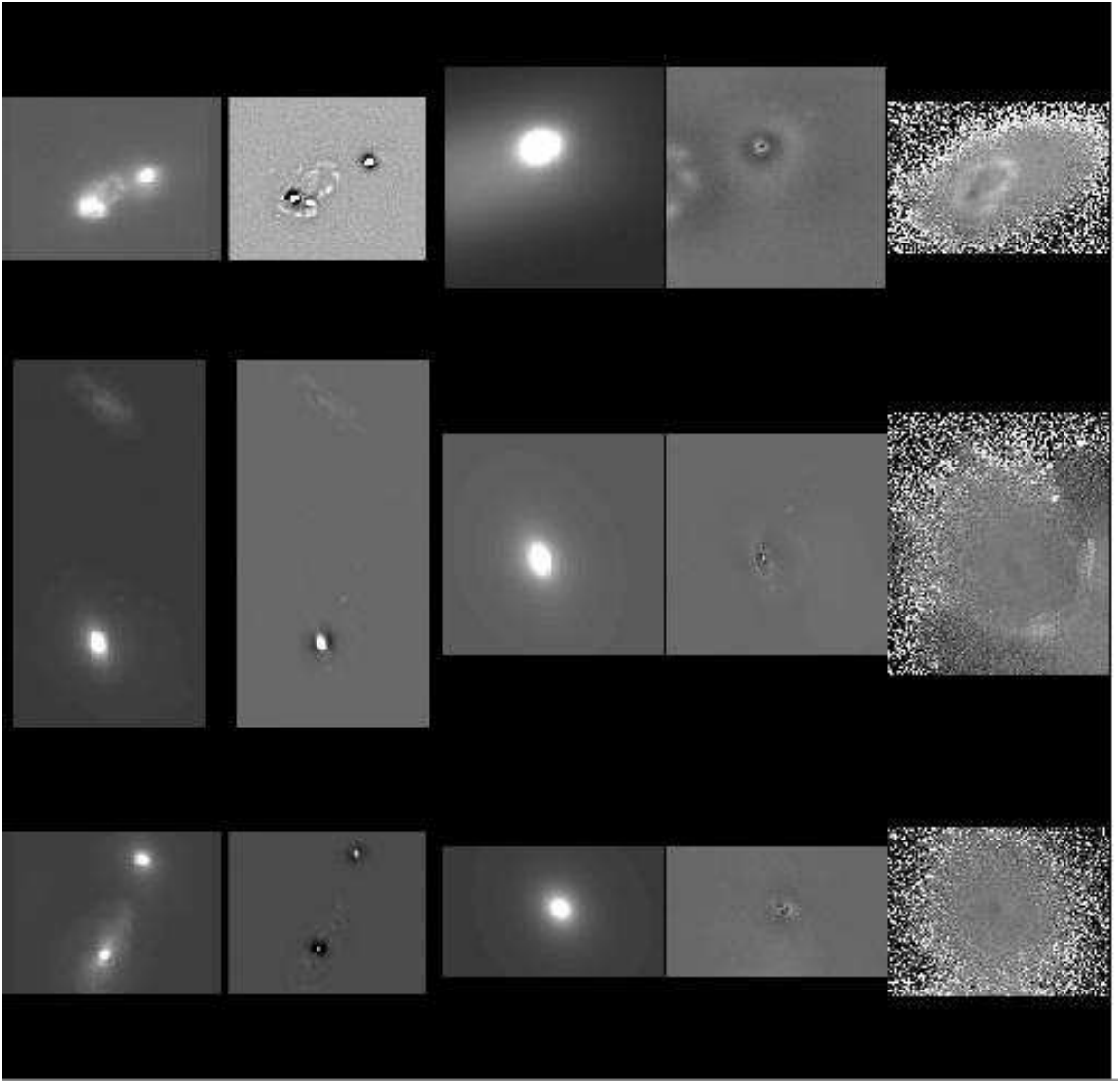}
\caption{A mosaic of images illustrating from left to right: An $R$ band image of a pair, the 
corresponding Unsharp Masking image, a GALFIT modell of the E/S0 component 
galaxy light distribution, a residual (original - model) image of the E/S0 component 
and a $B-I$ color index map. Upper panel: KPG 83, 
Middle panel: KPG 86 and Lower panel: KPG 234.
 \label{fig5}}
\end{figure*}

\subsection{$CAS$ parameters of bright E galaxy precursors}

On the ground of semi-analytic models and simulations, it was proposed 
that the local population of  ellipticals assembled a significant fraction of 
their masses via late major (``dry'') mergers (Kauffmann \& Haehnelt 2000;  
Khochfar \& Burkert 2003; Dom\'{\i}nguez-Tenreiro et al. 2005). Non-dissipative 
mergers seem to be the main mechanism to produce the present-day massive 
anisotropic, slowly rotating and boxy ellipticals (Naab et al. 2006). 
Recently, from an analysis of tidal features associated with bright red 
galaxies, van Dokkum (2005;  see also Tran et al. 2005) has found that 
$\sim 35\%$ of early-type galaxies experienced a ``dry'' 
major merger involving more than 20\% of its final mass. The E/S0's in mixed 
pairs analyzed here, at least those with clear signatures of interactions, 
may serve as local examples of the population of early-type galaxies
in a previous stage to a major merger at higher redshifts. Thus, according to 
our results, the $CAS$ criteria of pre-merger precursors of bright ellipticals 
should be $A>0.05$, and $0.86 -0.17C < S < 0.3$ (see above and Fig. 5).
We note that several of the interacting E/S0's have colors slightly bluer 
than the non-interacting E/S0's (\S 3.4 and Fig. 6).

Finally, in Fig. 9 we ressume how the classification of various galaxies is 
envisaged from the point of view of $CAS$ planes. The average 
and $1\sigma$ dispersion $CAS$ values of non-interacting isolated (crossed bars) and 
interacting (circled bars) galaxies in different morphological-type ranges are shown. 
For the interacting galaxies, we use the subsets of paired E/S0's (this paper), and 
paired SaSb's and ScSm's (from H2005) that obey the corresponding interaction 
criteria. 
For completness, we also plot the data for the 
ULIR, starburst, and dwarf E galaxy samples (dotted error bars, see C2003).

Different galactic environments (from mostly isolated to mergers) can be 
differentiated in this structural-morphological space. If the $CAS$ parameters can be
measured reliably at various S/N ratios, resolutions and redshifts, this 
diagram can be a useful tool for exploring different galaxy populations at 
larger redshifts.   
 
\begin{figure*}
\epsscale{2}
\plotone{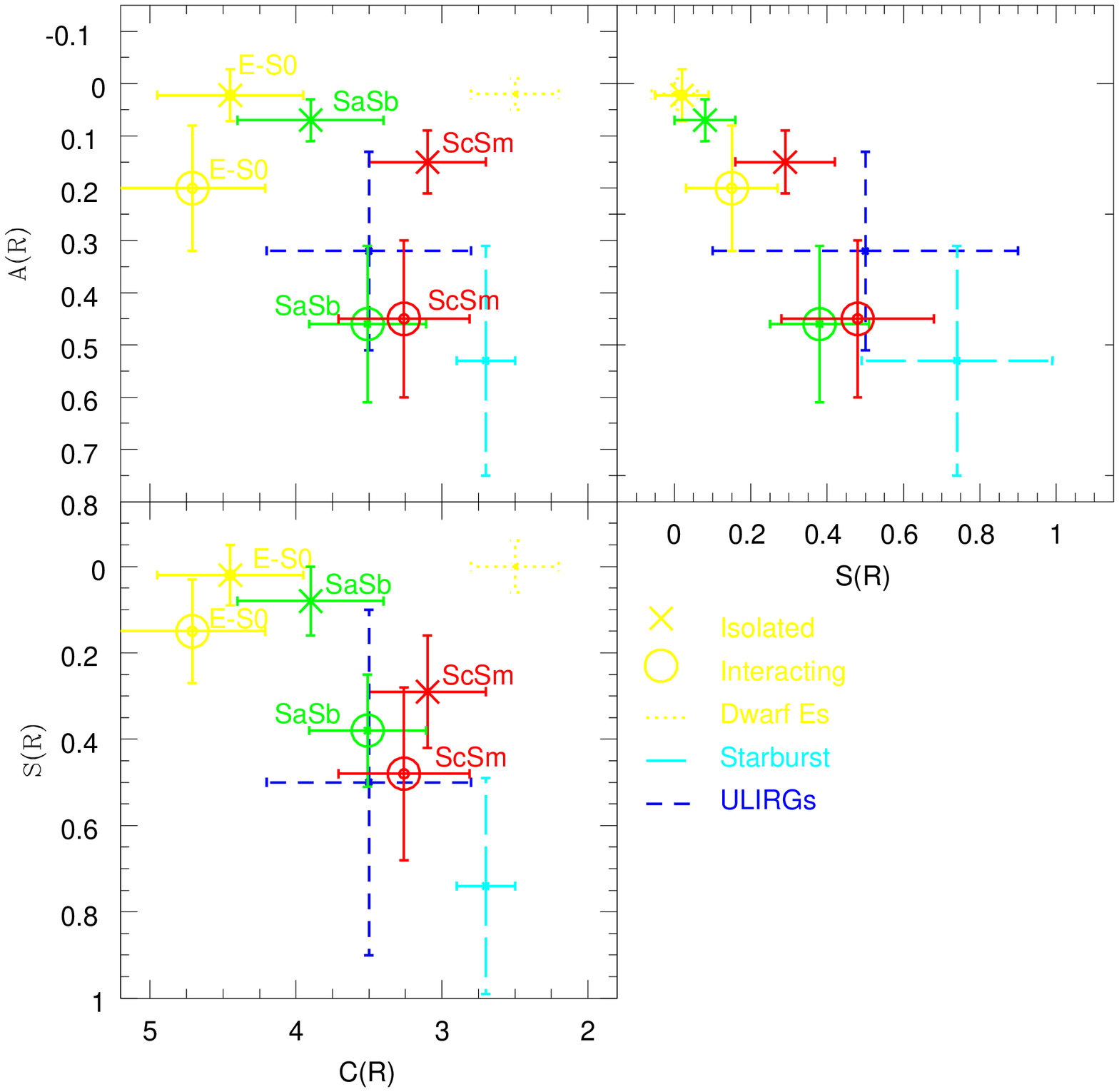}
\caption{The loci of different galaxies in the $CAS$ parameter space 
as a function of their morphological types. The average and $1\sigma$ dispersion $CAS$ 
values of non-interacting isolated (crossed bars) and interacting (circled bars) galaxies 
of different morphological-type ranges are shown. Paired E/S0's come from this paper 
while paired Sa-Sbs and Sc-Sms come from H2005. For completness, we also plot the data 
for the ULIR, starburst, and dwarf E galaxy samples (dotted error bars, see Conselice 2003).
 \label{fig5}}
\end{figure*}

\section{Conclusions} \label{S5}

We have analyzed 42 images in the optical $BVRI$ bands of E/S0 galaxies
in mixed morphology pairs selected from the Karachentsev (1972) catalogue.  
We measured concentration ($C$), asymmetry ($A$) and clumpiness ($S$) 
($CAS$) parameters on these images using the C2003 methodology. 
Our goal was to explore the effects of interactions on paired E/S0 
galaxies by comparing their structural parameters to those of mostly
isolated non-interacting E/S0's (our isolated sample comprissed 67 galaxies, 
half of them observed by us). Our main results are:

1. The difference in the means of the asymmetry parameter $A$ between
paired and non-interacting E/S0's is statistically significant, their ratio 
being $2.96\pm 0.72$. For the $25\%$ subset of galaxies with the highest 
asymmetries in each sample, $A$ can be several times larger for the paired E/S0's 
with respect to the non-interacting ones. The concentration parameter $C$ of both 
samples is statistically similar. For the subsets with the highest $C$ values in 
both samples, the $C$ values are slightly larger for the paired E/S0s than for the 
non-interacting ones. For the $S$ parameter, the the median $S$ value of the 
former is $\sim 2$ times larger than the median of the latter, but the difference 
is not statistically significant.

2. The paired E/S0 galaxies occupy a more scattered loci in $CAS$ space than
non-interacting E/S0's, but still quite narrow with respect to paired
disk galaxies. There is a subset of paired E/S0's (presumably the most
interacting ones) that deviate in the $CAS$ planes from the rest of paired 
E/S0 and non-interacting galaxies in the direction 
of higher values of $A$ and $S$. 

3. There are no significant correlations between the $CAS$ parameters
and magnitude, total $(B-V)$ color, central velocity dispersion, and 
20cm continuum luminosity for both paired and non-interacting E/S0's. 
The scatter in all of these diagrams are larger for the former than for the 
latter, but there are no systematic differences, with the exception of the 
$CAS$ vs $(B-V)$ color diagrams, where a subset of blue paired E/S0's clearly 
appears.   

4. Our results indicate that early-type galaxies in a pre-merging phase
would not be easily recognized through standard interaction/merger criteria 
that are suitable for gas rich galaxies. We find that a rough statistical 
criterion for finding presumably pre-merging E/S0 galaxies might be: 
$A>0.05$ (alternativley, $\sigma_{AS}>1$) and $0.86 -0.17C < S < 0.3$. 
For the E/S0's galaxies in our sample of mixed pairs that obey these criteria, 
an image analysis show indeed morphological evidence of interaction. A very rough 
estimate of the average dynamical friction (major merging) timescale of these pairs
is $\grtsim 0.1-0.5$ Gyr.

\noindent Our main conclusions from this study are:

$\bullet$ Interactions do not produce significant changes in the 
morphological/structural/SF properties of present-day E/S0 galaxies, 
suggesting that most E/S0 galaxies in pairs are dense, dynamically hot 
spheroids with low gas fractions and probably slow rotation.

$\bullet$ The main effect of  interactions on E/S0 galaxies 
are to make them moderatively asymmetric, probably through $m=1$ tidal 
deformations, and to increase the clumpiness for the most interacting 
cases. 

$\bullet$ The paired E/S0 galaxies analyzed here seem to span a large 
variety of dynamical interaction levels as well as gas effects, such 
as cross-fueled transfer.

$\bullet$ In the case the differences in the $CAS$ parameters between
pair and isolated E/S0's were dominated by differences in the formation
processes rather than by the interaction, our results show some
evidence against a formation scenario of E/S0s based only on the 
secular (enhanced by interactions) mechanism. 

$\bullet$ The loci of our strongly interacting E/S0's in the $CAS$ space
might correspond to the loci of (``dry'') pre-merger early-type 
precursors of present-day bright ellipticals. Thus, our results can serve
to characterize these pre-merger early-type population at higher
redshifts. 

\acknowledgements
We thank J. Colbert for sharing his $R$-band observations of 
20 truly isolated E/S0 galaxies. We acknowledge the anonymous 
referee for a careful reading of the manuscript and suggestions
that improved it.
Support for this work was provided by CONACyT grant 
42810/A-1 to H.H.T. Support from an NSF 
Astronomy and Astrophysics Fellowship is acknowledged by C.J.C.












\clearpage

\begin{deluxetable}{rrrrrrrr} 
\tablecolumns{8} 
\tablewidth{0pc} 
\tablecaption{$CAS$ Parameters for Paired and Isolated E/S0 Galaxies. \label{tbl-1}} 
\tablehead{ 
\colhead{}    &  \multicolumn{3}{c}{$R$-band} &   \colhead{}   & \\ 
\cline{2-5}\\   
\colhead{KPG} & \colhead{$M_{B}$}   & \colhead{$C$}    & \colhead{$A$} & 
\colhead{$S$}}
\startdata

KPG101b & -21.44&  4.14$\pm$0.15& 0.02$\pm$0.03& 0.03$\pm$0.09&  \\
KPG129a & -20.78&  4.93$\pm$0.23& 0.02$\pm$0.01& 0.08$\pm$0.09&  \\
KPG162a & -21.40&  4.22$\pm$0.19& 0.00$\pm$0.08& 0.02$\pm$0.17&  \\
KPG191a & -20.43&  3.94$\pm$0.58& 0.06$\pm$0.01& 0.07$\pm$0.18&  \\
KPG202a & -20.31&  4.46$\pm$0.16& 0.00$\pm$0.00& -0.02$\pm$0.08&  \\
KPG229a & -20.11&  4.84$\pm$0.42& 0.08$\pm$0.02& 0.07$\pm$0.23&  \\
KPG234a & -21.04&  4.35$\pm$0.10& 0.05$\pm$0.01& -0.02$\pm$0.13&  \\
KPG239a & -20.02&  4.43$\pm$0.43& 0.07$\pm$0.01& 0.11$\pm$0.07&  \\
KPG254b & -20.74&  4.56$\pm$0.37& 0.08$\pm$0.01& 0.03$\pm$0.15&  \\
KPG260b & -20.34&  4.54$\pm$0.20& 0.10$\pm$0.00& 0.21$\pm$0.11&  \\
KPG29b  & -20.66&  4.16$\pm$0.12& 0.01$\pm$0.02& -0.03$\pm$0.18&  \\
KPG303a & -21.28&  4.40$\pm$0.21& 0.04$\pm$0.01& -0.004$\pm$0.01&  \\
KPG339b & -20.66&  4.60$\pm$0.26& 0.03$\pm$0.02& -0.02$\pm$0.05&  \\
KPG353b & -21.13&  4.57$\pm$0.08& 0.05$\pm$0.00& -0.01$\pm$0.09&  \\
KPG363a & -20.28&  5.06$\pm$0.20& 0.22$\pm$0.01& 0.05$\pm$0.07&  \\
KPG386b & -19.77&  4.33$\pm$0.39& 0.03$\pm$0.02& -0.01$\pm$0.07&  \\
KPG38a  & -19.80&  4.40$\pm$0.16& 0.01$\pm$0.03& 0.01$\pm$0.07&  \\
KPG38b  & -19.80&  4.77$\pm$0.13& 0.01$\pm$0.04& -0.02$\pm$0.20&  \\
KPG392a & -19.84&  4.44$\pm$0.25& 0.06$\pm$0.01& 0.12$\pm$0.06&  \\
KPG393a & -21.02&  4.13$\pm$0.37& 0.07$\pm$0.01& 0.06$\pm$0.02&  \\
KPG394a& -21.54&  4.10$\pm$0.30& -0.04$\pm$0.04& -0.06$\pm$0.03&  \\
KPG407b& -21.58&  5.08$\pm$0.12& 0.49$\pm$0.00& 0.16$\pm$0.21&  \\
KPG408a& -21.08&  4.26$\pm$0.28& 0.01$\pm$0.00& -0.01$\pm$0.09&  \\
KPG416b& -20.46&  5.18$\pm$0.25& 0.01$\pm$0.02& 0.05$\pm$0.10&  \\
KPG419b& -20.15&  4.65$\pm$0.35& 0.04$\pm$0.01& -0.06$\pm$0.07&  \\
KPG429b& -20.42&  4.95$\pm$0.37& 0.02$\pm$0.01& 0.14$\pm$0.03&  \\
KPG432a& -21.25&  4.47$\pm$0.37& 0.00$\pm$0.01& -0.01$\pm$0.11&  \\
KPG445a& -21.03&  4.65$\pm$0.24& 0.01$\pm$0.01& 0.00$\pm$0.06&  \\
KPG460b& -20.79&  4.48$\pm$0.33& 0.16$\pm$0.01& 0.26$\pm$0.30&  \\
KPG487b& -21.13&  5.43$\pm$0.17& 0.24$\pm$0.03& 0.02$\pm$0.22&  \\
KPG526b& -19.68&  5.14$\pm$0.18& 0.03$\pm$0.01& -0.07$\pm$0.05&  \\
KPG542b& -19.70&  4.26$\pm$0.30& 0.18$\pm$0.00& 0.07$\pm$0.09&  \\
KPG548b& -20.51&  4.48$\pm$0.16& 0.02$\pm$0.01& 0.00$\pm$0.12&  \\
KPG552b& -20.12&  3.87$\pm$0.32& 0.04$\pm$0.00& 0.10$\pm$0.05&  \\
KPG553a& -19.28&  4.44$\pm$0.27& 0.04$\pm$0.02& 0.00$\pm$0.14&  \\
KPG572a& -20.17&  3.80$\pm$0.53& 0.06$\pm$0.02& 0.06$\pm$0.24&  \\
KPG591a& -17.85&  3.43$\pm$0.11& 0.20$\pm$0.01& 0.27$\pm$0.15&  \\
KPG61b& -19.61&  4.55$\pm$0.31& 0.02$\pm$0.01& -0.01$\pm$0.02&  \\
KPG62a& -19.26&  4.33$\pm$0.21& -0.00$\pm$0.02& 0.03$\pm$0.05&  \\
KPG81b& -20.86&  5.60$\pm$0.19& 0.13$\pm$0.05& 0.26$\pm$0.36&  \\
KPG83a& \nodata& 4.25$\pm$0.13& 0.19$\pm$0.01& 0.05$\pm$0.12&  \\
KPG86b& -18.34&  4.57$\pm$0.08& 0.07$\pm$0.02& 0.04$\pm$0.15&  \\
\enddata
\end{deluxetable}


\begin{deluxetable}{lccclcc} 
\tablecolumns{5} 
\tablewidth{0pc} 
\tablecaption{$R-$band Median) and quartiles of $CAS$ Parameters
for all E/S0's in Pairs, E/S0s in Pairs with $SEP<1$, and Non-interacting E/S0's}   
\tablehead{ 
\colhead{} &  & \colhead{(E+S) All/$SEP < 1$}  & & & \colhead{Frei+Colbert+Hdz-Toledo}  & \\
\colhead{}   & 75\% & 50\% (Median) & 25\% &  75\% & 50\% (Median) & 25\% }  
\startdata 

$C$ & 4.656/4.597 & 4.439/4.282 & 4.190/4.082 & 4.621 &  4.476 & 4.171 \\
$A$ & 0.067/0.150 & 0.035/0.047 & 0.009/0.015 & 0.039 &  0.019 & 0.007 \\
$S$ & 0.065/0.068 & 0.025/0.020 & -0.022/-0.018 & 0.057 & 0.014 & -0.029 \\
\enddata 
\end{deluxetable}

\end{document}